# A Cloud-based Real-time Probabilistic Remaining Useful Life (RUL) Estimation using the Sequential Monte Carlo (SMC) Method

Karthik Reddy Lyathakula and Fuh-Gwo Yuan
Department of Mechanical and Aerospace Engineering
North Carolina State University
Raleigh, NC 27695

**Abstract**

The remaining useful life (RUL) estimation is an important metric that helps in condition-based maintenance. Damage data obtained from the diagnostics techniques are often noisy and the RUL estimated from the data is less reliable. Estimating the probabilistic RUL by quantifying the uncertainty in the predictive model parameters using the noisy data increases confidence in the predicted values. Uncertainty quantification methods generate statistical samples for the model parameters, that represent the uncertainty, by evaluating the predictive model several times. The computational time for solving a physics-based predictive model is significant, which makes the statistical techniques to be computationally expensive. It is essential to reduce the computational time to estimate the RUL in a feasible time. In this work, real-time probabilistic RUL estimation is demonstrated in adhesively bonded joints using the Sequential Monte Carlo (SMC) sampling method and cloud-based computations. The SMC sampling method is an alternative to traditional MCMC methods, which enables generating the statistical parameter samples in parallel. The parallel computational capabilities of the SMC methods are exploited by running the SMC simulation on multiple cloud calls. This approach is demonstrated by estimating fatigue RUL in the adhesively bonded joint. The accuracy of probabilistic RUL estimated by SMC is validated by comparing it with RUL estimated by the MCMC and the experimental values. The SMC simulation is run on the cloud and the computational speedup of the SMC is demonstrated.

## 1. Introduction

Adhesively bonded joints are widely used in engineering applications, and it is vital to estimate their remaining useful life (RUL) [1-3]. Accurate estimation of RUL is a critical element of condition-based maintenance, as it helps to ensure efficient cost management and avoid catastrophic failure by enabling optimal scheduling of maintenance [4-8]. To estimate RUL, a predictive model is typically calibrated using damage data obtained from diagnostic techniques. However, the damage degradation data is often noisy, which introduces uncertainty in the RUL estimation. By quantifying this uncertainty, a probabilistic RUL estimate can be obtained, which enhances the confidence level of the estimation [9-11]. However, the techniques used to quantify uncertainty are computationally intensive [9, 12-15] and require methods to significantly reduce the computational time [16-19].

To accurately estimate RUL, it is crucial to have a predictive model that precisely describes damage growth in adhesive joints. There are several types of predictive models, including data-driven, physics-based, and hybrid models. Hybrid models, which combine physics-based and data-driven models, have demonstrated relatively high accuracy with significantly reduced computational time [9, 20-22]. This study uses a hybrid fatigue damage growth (FDG) simulator developed in recent works as the predictive model to simulate damage growth in adhesively bonded joints [13]. The FDG simulator integrates a data-driven artificial neural network (ANN) model for strain calculations with a physics-based damage growth model.

Calibrating the predictive model requires estimating model parameters using experimental damage degradation data, and quantifying the uncertainty in these parameters during model calibration requires the use of Bayesian inference [12]. Bayesian inference represents the quantified uncertainty in model parameters using a posterior distribution, which is a complex function. Statistical sampling methods, such as Markov Chain Monte Carlo (MCMC) and Sequential Monte Carlo (SMC), are used to approximate the posterior distribution. MCMC generates numerous parameter samples from the parameter posterior sequentially in a chain process, and the predictive model is solved for each of these parameter samples in the MCMC simulation [12, 23] . The computational time required for the MCMC simulation is mostly taken up by solving the predictive model, which is computationally intensive even with hybrid models. Alternatively, SMC methods can generate parameter samples using parallel computations, which significantly reduces computational time [16, 18, 19]. With increasing cloud computing techniques [24-26], parallel SMC methods can be exploited for real-time RUL estimation.

This study employs the SMC method and cloud computing to estimate real-time RUL in an adhesively bonded joint. The case study of mixed-mode failure (MMF) joint [27] is used to demonstrate the real-time RUL estimation. The first step involves quantifying and comparing the uncertainty in the model parameters of the FDG simulator between the MCMC and SMC methods. Then, the probabilistic RUL is estimated using samples generated from both the MCMC and SMC methods and compared with the experimental value. Further, the study showcases the scalability of the SMC methods by exploiting their parallelization capabilities and cloud computations.

The article is structured as floows: Section 2 provides a summary of the probabilistic framework utilized for estimating probabilistic RUL. A concise description of the predictive model is given in section 3. In section 4, the Bayesian inference for uncertainty quantification and sampling methods are presented. Lastly, the results are presented in section 5.

## 2. Remaining Useful Life Estimation Framework

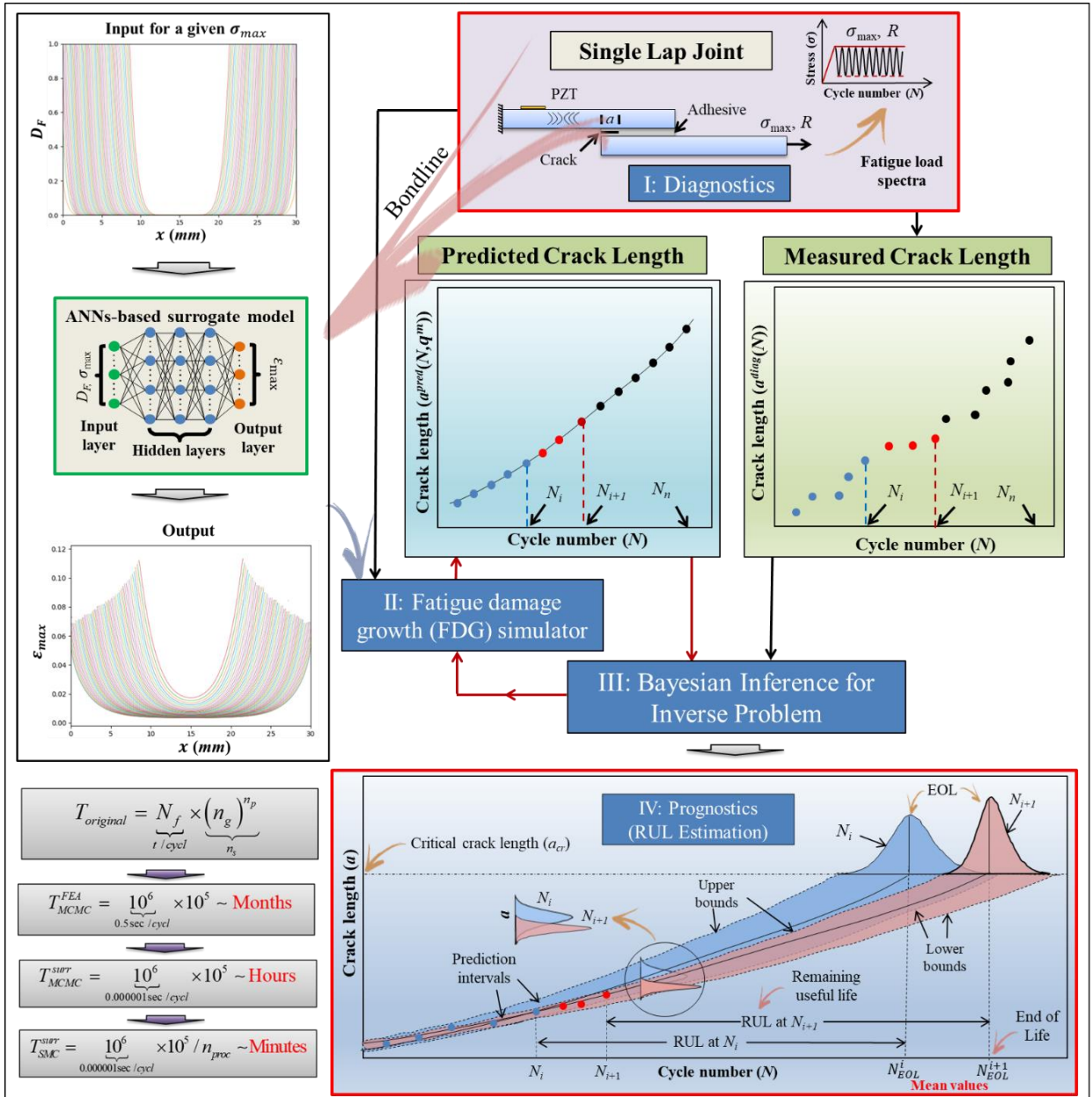

Figure 1. Schematic of the steps in real-time probabilistic RUL estimation in adhesive joints [10, 20].

Figure 1 depicts the schematic of the prognostics framework, adopted from [10], which is used to estimate probabilistic RUL in adhesive joints. The framework consists of four parts, namely: (1) diagnostics techniques to obtain the crack length, (2) development of the predictive model, which is the fatigue damage growth simulator, (3) Bayesian inference to quantify the uncertainty in the model parameters of the simulator and (4) probabilistic RUL estimation from the quantified model parameter uncertainties. The fatigue damage growth simulator incorporates a data-driven artificial neural network (ANN) model for calculating strains in the adhesive region, along with a physics-based damage growth model. Further details regarding the FDG simulator are discussed in section 3.

The Bayesian inference is used to estimate uncertainty by generating parameter samples from the posterior distribution using either MCMC or SMC methods. These sampling methods require the predictive model to be solved several times. Using the FEA model as the predictive model with MCMC takes months to quantify uncertainty. However, if the hybrid FDG simulator is used as the predictive model with MCMC, the computational time is reduced to hours. By utilizing parallel SMC methods with the hybrid FDG simulator, the computational time for uncertainty quantification simulation will be minutes. Sections 3 and 4 provide more information on each of the components presented in Figure 1.

## 2.1. Predictive Model

In the UQ step of the prognostic framework, the predictive model takes most of the computational time. The predictive model needs to be accurate with a reasonable amount of simulation time. Finite element methods have been used for computing the damage in mechanical structes [42,43] however the methods are computationally intensive. Previous studies [21, 22, 28-31] have shown that surrogate modeling can significantly reduce computational time. In this work, fatigue damage growth (FDG) simulator is used as the predictive model [14, 21]. This simulator reduces computational time by three orders while maintaining accuracy, as detailed in the reference [13]. The FDG simulator integrates a data-driven artificial neural network (ANN) model and a physics-based fatigue damage evolution law, as illustrated in Figure 2. The fatigue damage growth simulator is the predictive model [13, 20]. The adhesive is assumed to be under cohesive failure, and the progressive damage in the adhesive is modeled using the cohesive zone model (CZM) with a bilinear traction separation law. The strain field in the adhesive is calculated by the traction separation law, which is computationally expensive. To mitigate this, the ANN model is used as a surrogate model to replace the strain field calculations. The fatigue damage evolution law simulates the deleterious effect of fatigue by degrading the traction separation law. The ANN model is trained using data obtained from finite element simulations. The details of the failure modeling in adhesive, the FE model, and the training, testing, and validation of the FDG simulator are discussed in the references [13, 20, 27, 32-36].

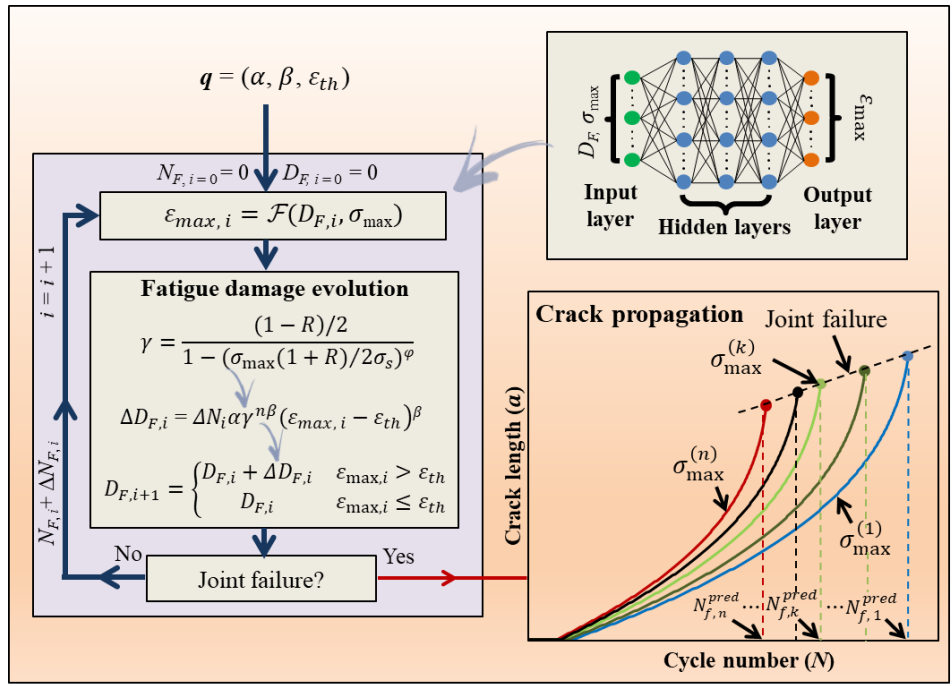

Figure 2. The fatigue damage growth simulator is the predictive model [13, 20].

### 2.2. Bayesian Inference

The Bayesian inference starts by defining the relationship between the experimental data and predicted values using the statistical model as shown below

$$a^{exp} = a^{pred}(\boldsymbol{q}) + \varepsilon_k \quad k=1,\cdots,n \tag{1}$$

where $a^{exp}$ is the random variable (RV) which represents the experimental crack length and RV $a^{pred}(\boldsymbol{q})$ is the computationally estimated value using the FDG simulator. The RV $a$ is the crack length data at different fatigue cycle numbers. $q$ in the statistical model is the RV that represents the model parameters of the simulator. The $\varepsilon_k$ is the error between the data and the predicted value. The error is assumed to be *iid* and normally distributed with a standard deviation, $\sigma$. $n$ is the number of data points and $k$ is the index of the data point.

In the statistical model, the $a^{exp}$ is the experimental crack length data and $a^{exp}(\boldsymbol{q})$ is obtained from the FDG simulator. From the statistical model, the Bayes theorem for inverse problems [12] estimates the model parameter uncertainties through the posterior distribution given by

$$\pi(\boldsymbol{q}|\boldsymbol{a}^{exp}) = \frac{\pi(\boldsymbol{a}^{exp}|\boldsymbol{q})\pi_0(\boldsymbol{q})}{\pi(\boldsymbol{a}^{exp})} = \frac{\pi(\boldsymbol{a}^{exp}|\boldsymbol{q})\pi_0(\boldsymbol{q})}{\int_q \pi(\boldsymbol{a}^{exp}|\boldsymbol{q})\pi_0(\boldsymbol{q})d\boldsymbol{q}} \tag{2}$$

where $\pi_0(\boldsymbol{q})$ is the prior, $\pi(\boldsymbol{q}|\boldsymbol{y}^{exp})$ is the posterior distributions respectively, and $\boldsymbol{a}^{exp} = (a_1^{exp},\cdots,a_n^{exp})$ are the experimental data points. From the assumption of errors ($\varepsilon_k$) are normally distributed, the likelihood function can be expressed as

$$\pi(\boldsymbol{y}^{exp}|\boldsymbol{q}) = \frac{1}{(2\pi\sigma_{sd}^2)^{n/2}} e^{-SS_q/2\sigma_{sd}^2} \tag{3}$$

where $SS_q$ is the sum of squares error which is given by

$$SS_q = \sum_{i=1}^n \left(a_i^{exp} - a_i^{pred}\right)^2 \tag{4}$$

The posterior distribution given by Eq. (2) represents the quantified parameter uncertainties.

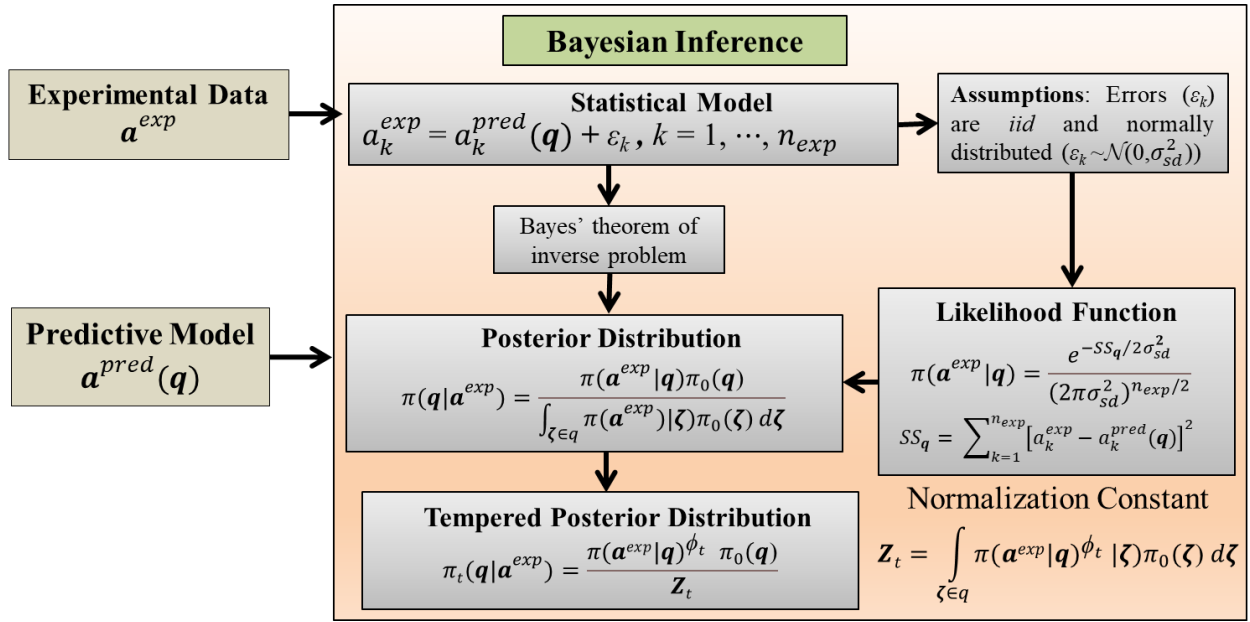

Figure 3. Bayesian inference for quantifying uncertainty in the model parameter.

Figure 3 shows the summary of the Bayesian inference for quantifying model parameter uncertainty presented in this section. The statistical model assumes that the error between the experimentally determined crack length and the estimated by the predictive model is independent and identically (*iid*) distributed with the errors assumed to be normally distributed. The Bayes theorem of the inverse problem represents the parameter posterior distribution as a function of likelihood and prior distribution. The likelihood function is obtained from the assumptions of the errors and the normalization constant is the integration of the likelihood and prior. To quantify the uncertainty, the normalization constant in the posterior needs to be solved. The parameter posterior is a highly nonlinear equation and is difficult to estimate using standard numerical techniques. The sampling methods are used to efficiently generate statical parameter samples that approximate the posterior. In this work, the Markov Chain Monte Carlo method and sequential Monte Carlo methods are used for generating samples from the posterior. The computational performance of the methods is compared.

### 2.2.1. Markov Chain Monte Carlo

The Markov chain Monte Carlo (MCMC) method is a statistical technique to approximate highly nonlinear posterior distribution by generating parameter samples from the posterior. The MCMC algorithm is based on the Markov chain principle and it generates parameter samples one after the other in a chain process, where one sample is generated only based on the previous sample. Figure 4 shows an overview of the MCMC algorithm to generate parameter samples from the posterior shown in Eq. (2)

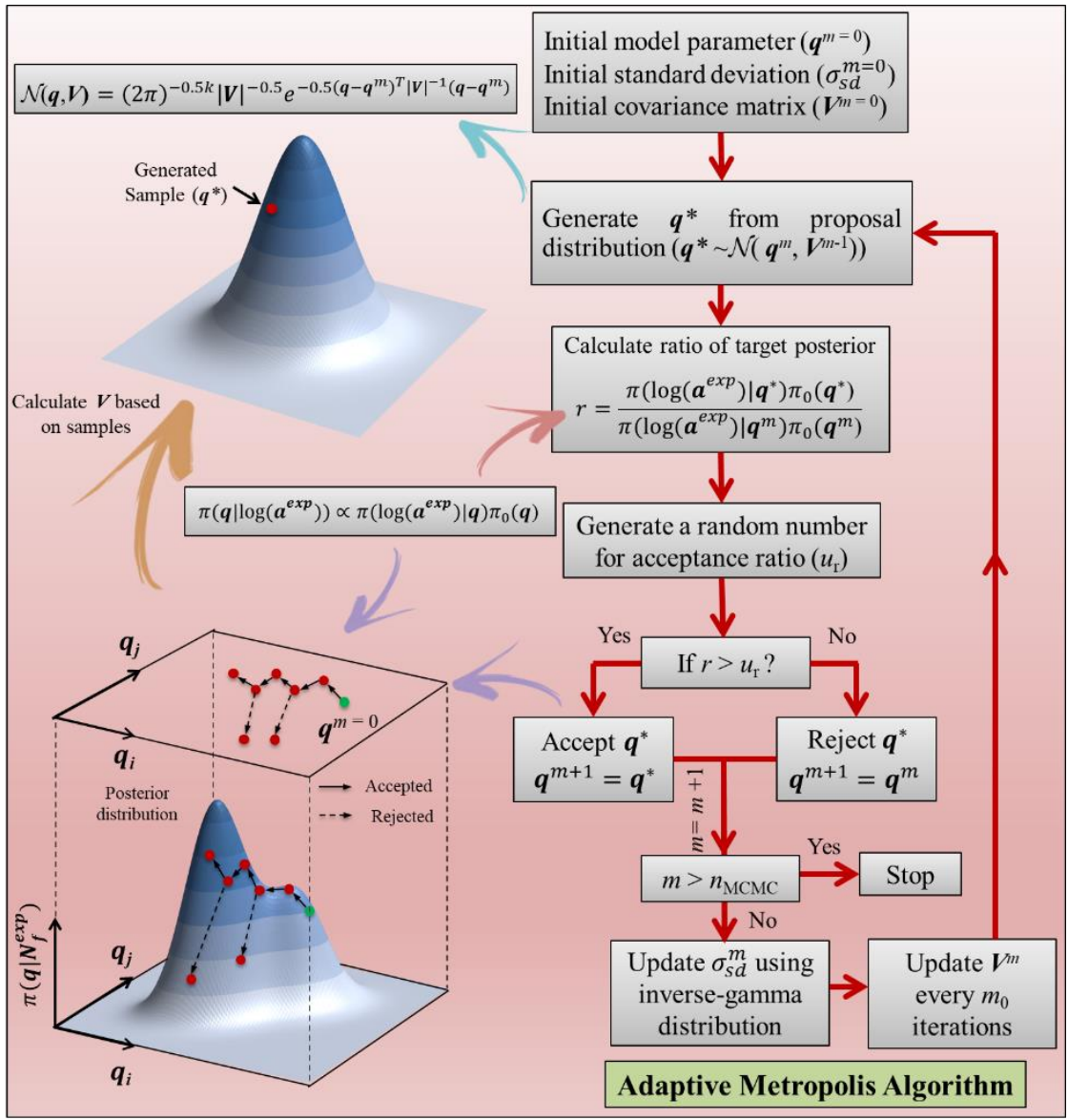

Figure 4. Adaptive Metropolis algorithm for generating samples from the parameter posterior [12, 13, 23].

The MCMC algorithm starts from an initial model parameter, error standard deviation, and parameter covariance matrix. First a parameter sample $q^*$ is generated from the normal distribution with the previous parameter sample as the mean and with a parameter covariance. Next, the ratio of the posterior for the parameter sample $q^*$ and for the previous parameter $q$ is evaluated. Taking the ratio avoids evaluating the complex normalization constant shown in Figure 3. In the next step, a random number is generated from the uniform distribution (ur) which determines accepting the generated parameter $q^*$. If the ratio $r$ is greater than $u_r$, the $q^*$ is accepted as the new parameter or else it is rejected. This chain process is repeated a large number of times to approximate the posterior. More details of the algorithm are explained in the references.

During the step of calculating the ratio of the posterior, the likelihood is evaluated, which is the function of the sum of squares between the crack predicted by the predictive model and the experimental data. The predictive model in this study is the fatigue damage growth simulation, which takes most of the computational time. The flowchart presented in Figure 4 shows that the MCMC sampling methods are inherently serial, which limits the exploitation of the computational speed-up provided by the FDG simulator and hinders real-time life predictions. The SMC method quantifies the uncertainties by parallelizing the sampling process and significantly reducing computational time. In the next section, the details of the SMC methods are presented.

### 2.2.2. Sequential Monte Carlo (SMC) Method

The sequential Monte Carlo (SMC) method is based on the particle filter and uses importance sampling and resampling techniques. Figure 5 shows the overview of the SMC method [16] and the complete details of the SMC method are described in the references. Here, a brief overview of the SMC method is discussed for completeness, and the details of the adoption of the method for cloud computations. The SMC method generates the samples through a sequence of temperature distribution. The tempered distribution is controlled by $\phi$ and the initial condition for $\phi$ is 0. The method starts by randomly generating the samples from the prior and each sample is a particle that is moved around the parametric space in search of the posterior distribution. Each particle is associated with a weight indicating how close is the sample from the mode of posterior and an equal weight is assigned to each particle at the start of the simulation. The weights of the particles are adjusted based on the tempered distribution and the particles are mutated to explore the tempered posterior distribution using the MCMC kernel. In the next step, the $\phi$ of the tempered distribution is incremented and $\Delta\phi$ increment is obtained using adaptive methods by optimizing the effective sampling size (ESS) [20, 37]. The simulation is conducted until the $\phi = 1$.

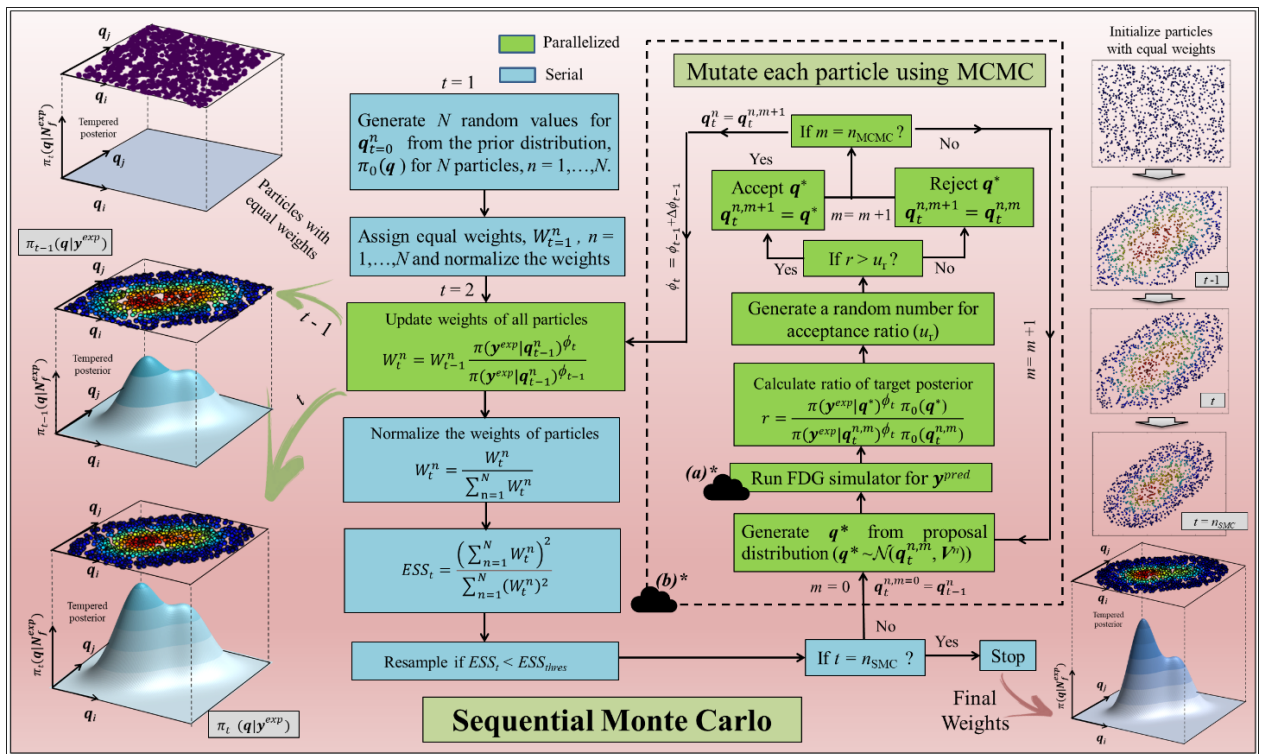

Figure 5. Sequential Monte Carlo method for generating parameter samples from the posterior distribution adopted from [16, 18-20] for cloud-based remaining useful life (RUL) estimation.

The SMC method can be run in parallel and the green boxes in Figure 5 show the part of the method that can be run in parallel. In the mutation part, the predictive model is solved multiple times for each particle. The mutation part of the SMC method takes most of the computational time and parallelization of the mutation gives significant speedup. Further to parallel computations of the mutation part, vectorization gives significant speedup. In the recent article, the speedup of the vectorized and parallel computations of the SMC method on-premises cluster was demonstrated. In this article, the parallel computation of the SMC on the cloud is demonstrated. For the cloud computations, the non-parallel part of the method is conducted on the local computer and the mutation part is computed on the cloud architecture using API calls. The SMC simulation is conducted in a serial computation on a local computer with Intel® Core™ i7-9750H CPU @ 2.60GHz × 12 processors, 16 GB memory, and the multithreading is used to compute the mutation part by calling the cloud APIs. Cloud calls mostly involve sending a request and waiting for the reply to the request. Multithreading will serve the purpose. For the computations of the mutation, either the predictive model or the complete mutation kernel for each particle can be solved on the cloud. The speedup using both methods is demonstrated.

## 3. Results

This section presents the results of remaining useful life estimation using the SMC method and speedup of the SMC method.

### 3.1 Remaining Useful Life (RUL) Estimation

Figure 6a shows the geometry of the mixed mode flexural joint (MMF) used in this study and Figure 6b shows the material properties of the adhesively bonded joint. The MMF joint is placed on a roller support and a fatigue load is applied as a point load on the top of the adhesive-bonded joint. The crack starts propagating from the left end of the adhesive region and the joint is considered failed when the crack length reaches 20 mm. In this article, the experimental data conducted with fatigue spectrum with a load ratio of 0.1 and a maximum load of 50% of static fatigue failure load is used. More details of the experimental setup are discussed in the literature.

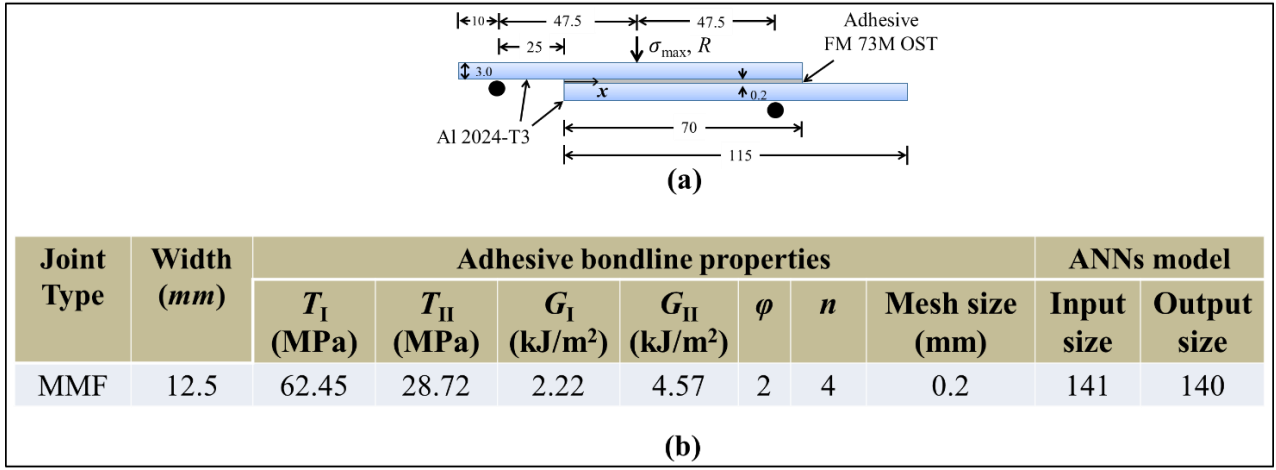

Figure 6. The details of the adhesively bonded joint for demonstrating cloud-based remaining useful life (RUL) estimation. (a) The geometry of the adhesively bonded joint and (b) the material properties of the adhesive bondline.

The FDG simulator for the MMF joint is developed before conducting the SMC simulation. To develop the FDG simulator, first FEA simulations are conducted for different values of fatigue model parameters, and the data obtained from the simulations are used for training the ANNs part of the FDG simulator. The FDG simulator with the training ANNs is validated and the results showed a speedup of three orders of magnitude with an error of less than 1%. More details of the training, testing, and validation of the FDG simulator for the MMF joint are discussed in the reference. Next, the developed FDG simulator is used for quantifying uncertainty in the model parameters of the FDG simulator using the experimental data and for probabilistic RUL estimation. Figure 7 shows the scatter plot experimental crack data under fatigue loading with a load ratio of 0.1, a maximum load of 50% of static failure load. The continuous line in the plot is the least square fit of the experimental data by the FDG simulator.

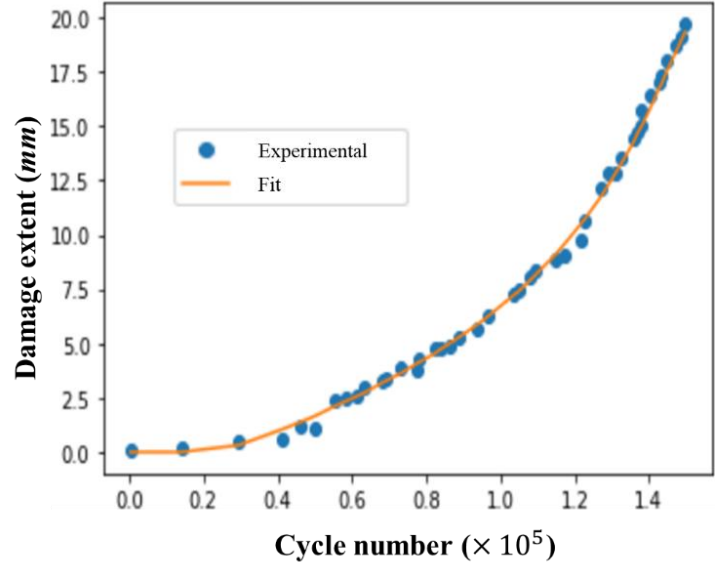

Figure 7. The experimental crack length data under fatigue loading and the least squares fit were obtained from the FDG simulator.

The experimental damage propagation data in the adhesively bonded joint is used for quantifying uncertainty in the model parameters of the FDG simulator. The MCMC and SMC are used for quantifying uncertainty by generating samples from the posterior distribution. In the MCMC method, the samples from the posterior distribution are generated in a serial process starting from an initial parameter $q^{m=0}$. The maximum-a-posterior (MAP) is used as the initial parameter. A total of 50,000 samples are generated and the initial 20,000 samples are removed as burn-in. The predictive model is evaluated for each generated parameter sample. In the SMC method, $N=1000$ particles are used to explore the parameter space to approximate the posterior with $n_{MCMC} = 5$ mutations in each step of the method. An adaptive approach is used to calculate the $\Delta\phi$ and the SMC simulation is terminated when the $\phi$ reaches a value of 1. Figure 8 shows the comparison of the pairwise plots of samples ($log_{10}(\alpha)$, $\beta$, $\sigma$) generated by the MCMC and SMC methods. The pairwise plots have a similar trend by MCMC and SMC methods with similar parameter means.

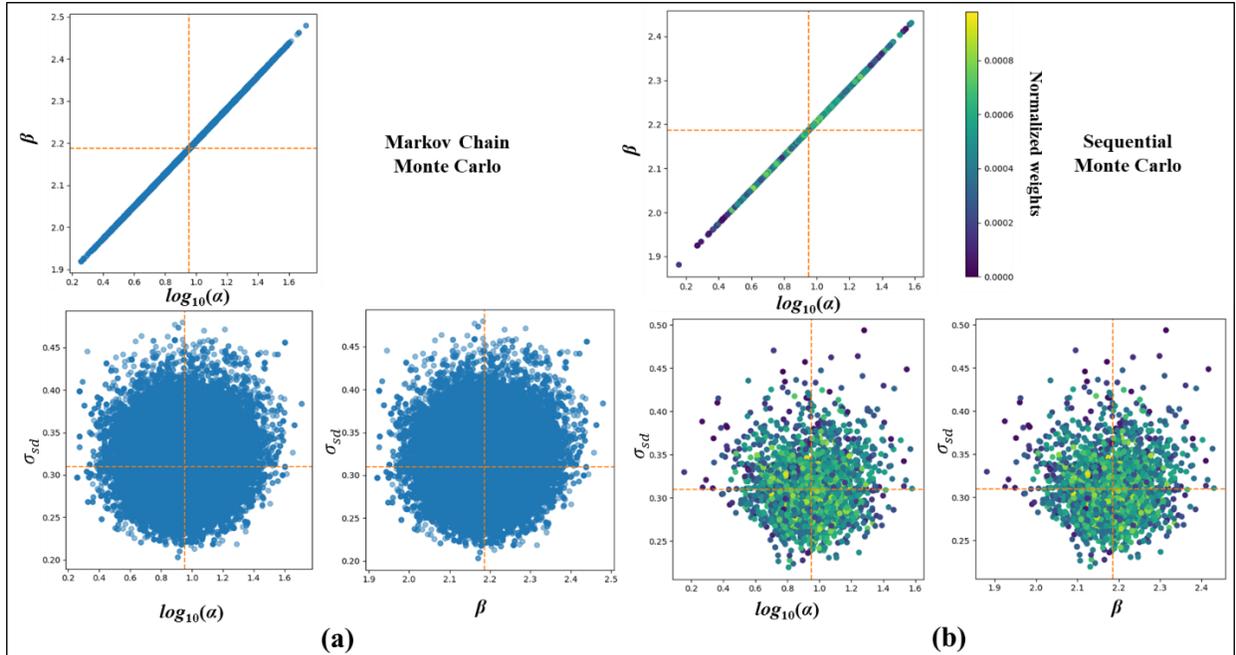

Figure 8. Pairwise plots obtained from uncertainty quantification using (a) MCMC method [10] and (b) SMC method.

The parameter samples obtained from the sampling method approximate the posterior distribution, which represents the uncertainty in the model parameters. From these parameter samples, probabilistic RUL is estimated. The parameter samples are propagated through the FDG simulator to calculate the predictive intervals in the crack propagation. To calculate the predictive intervals, first, the FDG simulator is solved for each parameter sample, and crack propagation is obtained. Next, the error values are generated for each cycle number from a normal distribution with the corresponding standard deviation value and added to the crack propagation. Finally, the crack length values are arranged in increasing order at each cycle number, and values corresponding to 95% confidence intervals are obtained. Figure 9a shows the uncertainty in crack propagation constructed using the parameter samples and Figure 9b shows the corresponding RUL. The RUL box plots are constructued by taking difference between the cycle number at which the crack length reaches failure crack length and the current cycle number.

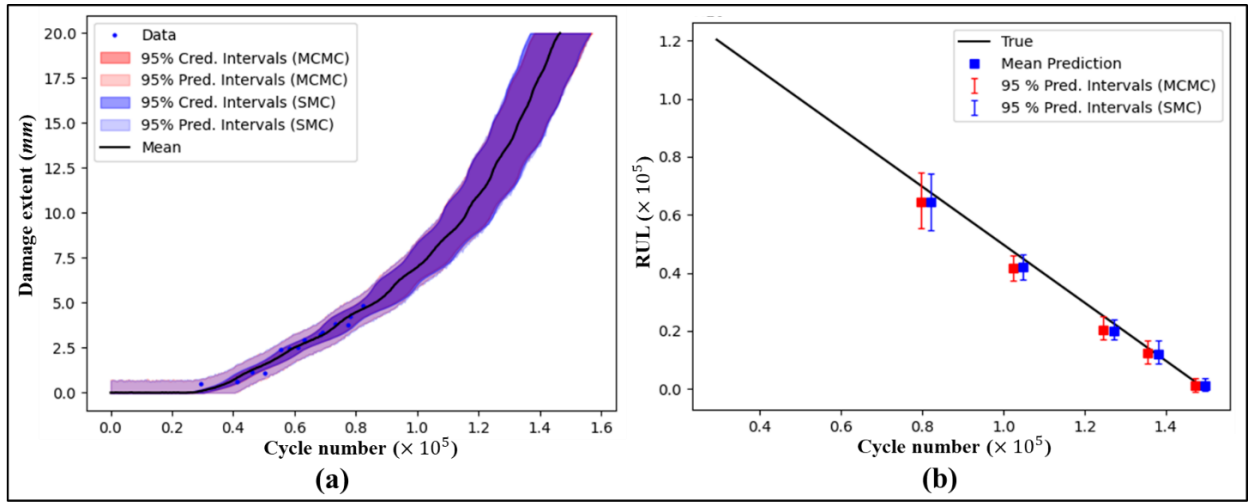

Figure 9. (a) Comparison of uncertainty in the crack propagation and (b) corresponding RUL using MCMC and SMC method. The SMC simulations are conducted with 1024 particles.

For the MCMC methods, a total of 50000 samples are generated with 20000 used as burn-in, and out of the remaining 30000, thining is conducted by picking every 5$^{th}$ sample in the chain to select 6000 samples. The UQ plots are constructed using the 6000 samples for the MCMC method. Figure 9a shows the comparison of uncertainty in crack propagation between the MCMC with resultant 6000 samples and SMC methods with 1024 particles. Figure 9b shows the corresponding RUL estimated from the uncertainty intervals. The figures show that the crack propagation intervals estimated using the SMC method overlap with the MCMC method. Further, the estimated RUL by the SMC method is close to the true RUL and similar to the MCMC method.

### 3.2. RUL Estimation on Cloud Platform

In the previous section, it is showed the probabilistic RUL estimated by the SMC method is similar to the values estimated by the conventional MCMC method. However, the SMC method can be run in parallel whereas the MCMC method requires serial computations. The parallel nature of the SMC method enables to use of the cloud computational resources [38-41] for rapid uncertainty quantification and enables real-time remaining useful life estimation. This section presents the results of the speedup of uncertainty quantification using the parallel SMC method compared to the MCMC method on cloud resources. The MCMC method is a serial algorithm and MCMC simulation is conducted on-prem machine.

The SMC method enables parallel computations, however only certain steps in the SMC method can be run on the parallel processes as shown in Figure 5. The mutation step, which can be run in parallel, takes most of the computational time with predictive model evaluations contributing to 99% of the simulation time of the SMC method. In the mutation step, the MCMC algorithm for each particle is independently simulated for $n_{MCMC}$ steps, and in each MCMC step, the predictive model (FDG simulation) is evaluated. These independent MCMC simulations along with the predictive models can be run in parallel to decrease the computational time and vectorized simulations can be used to further decrease the computational time. In this work, the vectorized SMC simulations are conducted on the cloud and computational performance calculations are presented. For the cloud computations, the serial steps of the SMC simulation are computed on-prem machine, and the parallelizable steps are conducted on the cloud through API calls. There are two methods for running the parallel steps on the cloud: (a) parallel computations of the FDG simulator on the cloud and (b) parallel computations of the mutations on the cloud as shown in Figure 5.

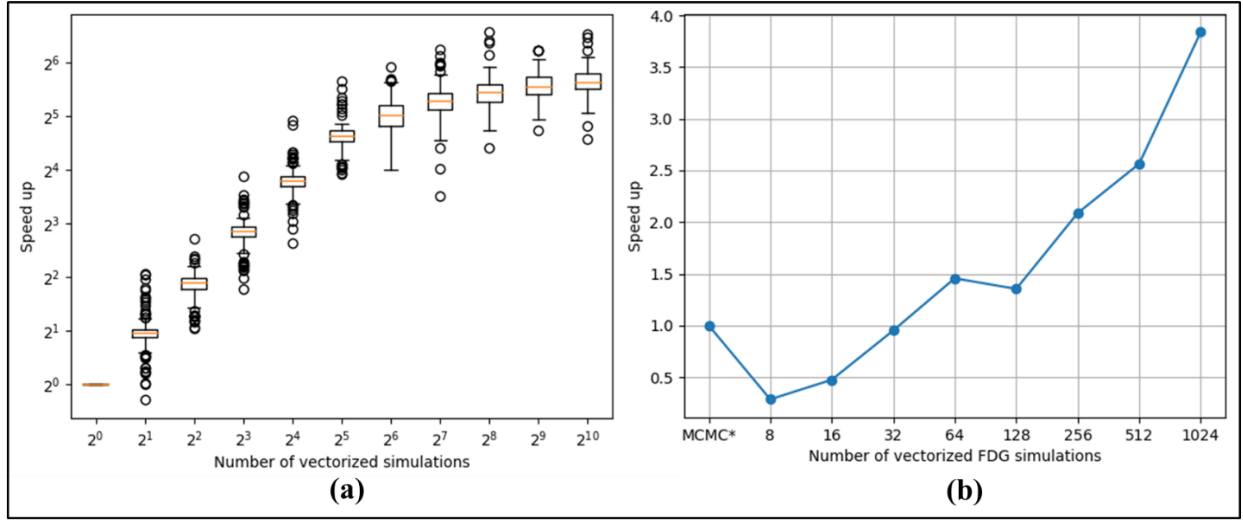

Figure 10. (a) Speedup of individual vectorized FDG simulations and (b) Speedup of SMC computations with vectorized FDG simulations.

First, vectorized FDG simulations are conducted on the cloud and the speedup of the simulations are calculated. Figure 10a shows the boxplots of computational speedup of independent vectorized FDG simulations on the cloud. To generate the plot, 100 independent simulations are conducted with different numbers of vectorized simulations. The results show that the median speedup increases with the number of vectorized simulations initially and reaches a plateau at 64 vectorized simulations. Next, the SMC simulations are conducted on the cloud with vectorized mutations. Figure 10b shows the variation in the speedup of serial SMC simulations, with vectorized FDG simulations, compared to serial MCMC simulations. The MCMC simulation is conducted on a local desktop computer with Intel® Core™ i7-9750H CPU @ 2.60GHz × 12 cores, 16 GB memory. The cloud machines used for MCMC mutations in the SMC method is a single core, 4 GB memory. The results show that speedup decreases first and then increases with the number of vectorized mutations compared to the MCMC simulation. This demonstrates that just vectorization shows four times speed up but which is not significant.

Next, the SMC simulations are conducted with parallel and vectorized FDG simulations. The mutation part in the SMC simulations takes most of the computational time with FDG simulations taking 99% of the computational time. The SMC simulations are conducted on-premises computer with the computationally intensive mutations on the cloud, as shown in Figure 5, through API calls. There are two ways to run the mutations on cloud: (a) Evaluating FDG simulations on the cloud (b) Evaluating entire mutations on the cloud. As the expensive mutations are conducted on the cloud, the non-parallelizable part is run in serial process and multithreading is used to call the API calls to run the parallel computations.

**(a) Evaluating FDG simulations on the cloud**

Figure 11a shows the speedup of vectorized FDG computations conducted on the cloud with number of threads. For the SMC simulation, 1024 particles ($N$) are used with 5 MCMC mutations ($n_{MCMC}$) in each SMC step ($n_{SMC}$). In each thread, the number of vectorized simulations are given by $N/n_{threads}$. In Figure 11a, the primary x-axis shows the number of threads and secondary x-axis is the number of vectorized FDG simulations. The MCMC simulation generates samples in a chain and requires evaluating the FDG simulation for each sample in a serial process with number of vectorized FDG simulations as one. In case of the SMC simulation, the FDG simulations for all particles (sample) are evaluated in vectorized and in parallel processes. The speedup is compared to the MCMC simulation on premises computer and the speedup increases with number of threads. It showed a maximum speedup of 14 with 32 number of threads and 32 number of vectorized simulations ($n_{threads} * n_{vectorization} = n_{particles}$).

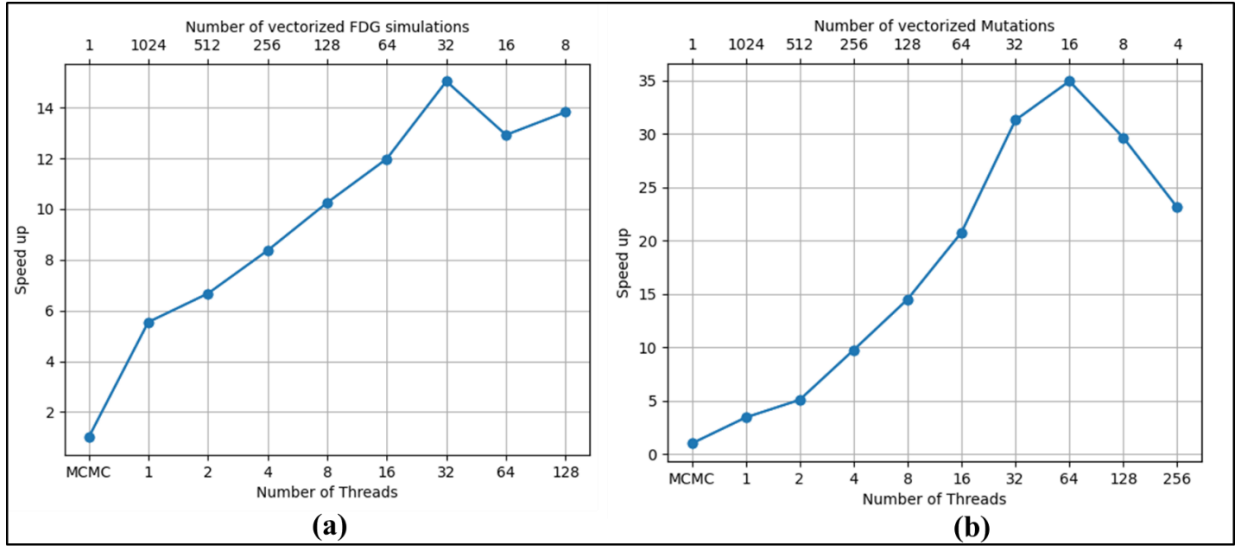

Figure 11. Speedup of SMC simulations on the cloud with multithreading and vectorized crack propagation simulations.

**(b) Evaluating entire mutations on the cloud**

In the second method, the MCMC mutations for all the particles in the SMC method are evaluated on the cloud. The MCMC mutation for each particle is independent and vectorization can be used to run MCMC mutations on multiple particles simultaneously. Figure 11b shows the speedup with the number of threads using the second method. The primary x-axis shows the number of threads and the secondary x-axis shows the number of vectorized simulations. The speedup with the second method is higher than the first method. The reason is the total number of API calls with the first method is higher than the first method and communication time increases with the total number of calls. The results show a speedup of 35 with 64 threads and 16 vectorized simulations on each thread. Finally, the computational time for generating samples from posterior distribution is reduced from 180 minutes using MCMC simulation on-premises computer to 5 minutes using SMC simulations on the cloud.

## 4. Summary and Conclusion

In this work, a framework for cloud-based RUL estimation in adhesively bonded joints is demonstrated using the SMC method and hybrid surrogate model. The estimation of RUL plays a crucial role in optimizing maintenance scheduling for condition-based maintenance applications. The noise in the data limits the accurate estimation of the RUL. Quantifying uncertainty in the noise and predicting probabilistic RUL increases confidence in estimations. Conventional Markov chain Monte Carlo (MCMC) sampling methods are commonly used for uncertainty quantification. However, these methods are inherently serial and require a significant amount of computational time, making them impractical for large-scale applications. There are parallel sampling methods such as Sequential Monte Carlo (SMC) method that can reduce computational time significantly. However, the speedup achieved by the SMC method is limited by the computational resources. In this work, a framework for cloud-based RUL estimation in adhesively bonded joints is demonstrated using the SMC method and hybrid surrogate model.

The fatigue damage growth (FDG) simulator is used as a predictive model to simulate the damage degradation in the adhesively bonded joints. First, the FDG simulator is calibrated by quantifying the uncertainty in the model parameters using the experimental crack length. Next, the quantified uncertainties are propagated back through the FDG simulator to estimate the probabilistic RUL. The uncertainty in the model parameters was quantified using both MCMC and SMC methods. Both methods estimated similar uncertainty in the crack length propagation and probabilistic RUL. The parallel nature of SMC methods has reduced computational time significantly compared to the MCMC method. By using the vectorized simulations, the computational time with SMC methods can be further reduced. The scalability of the SMC method is demonstrated by conducting the simulation on the cloud. Two approaches are used for computing the parallel steps on the cloud:

parallel computations of the predictive model and parallel computations of mutations in the SMC method. The parallel computations of mutations have shown a superior speedup with computational time for estimating probabilistic RUL reducing to 5 minutes. In conclusion, this work demonstrates that the SMC method shows computational advantages due to its parallel nature. Leveraging the cloud architecture reduces the computational time of the SMC method significantly and enables real-time RUL estimation.